\documentclass[doublecol,figures]{epl2}
\usepackage{amsmath,graphicx}
\usepackage[squaren]{SIunits}

\renewcommand{\vec}[1]{\mathbf{#1}}

\title{Quantum Preferred Frame: Does It Really Exist?} 

\author{J. Rembieli\'nski\thanks{E-mail: \email{jaremb@uni.lodz.pl}}
  \and K. A. Smoli\'nski\thanks{E-mail:
    \email{K.A.Smolinski@merlin.phys.uni.lodz.pl}}}

\shortauthor{J. Rembieli\'nski and K. A. Smoli\'nski}

\institute{Department of Theoretical Physics, University of Lodz,
  ul.\ Pomorska 149/153, 90-236 \L\'od\'z, Poland, EU}

\date{\today}


\abstract{The idea of the preferred frame as a remedy for difficulties
  of the relativistic quantum mechanics in description of the non-local
  quantum phenomena was undertaken by such physicists as J.~S. Bell
  and D.~Bohm.  The possibility of the existence of preferred frame
  was also seriously treated by P.~A.~M. Dirac.  In this paper, we
  propose an Einstein-Podolsky-Rosen-type experiment for testing the
  possible existence of a quantum preferred frame.  Our analysis
  suggests that to verify whether a preferred frame of reference in
  the quantum world exists it is enough to perform an EPR type
  experiment with pair of observers staying in the same inertial frame
  and with use of the massive EPR pair of spin one-half or spin one
  particles.}

\begin{document}

\maketitle

\section{Introduction}
\label{sec:intro}

Quantum mechanics (QM) is not a local realistic theory.  It predicts a
strange phenomenon, named by Einstein as ``spooky action at a
distance''.  It was confirmed in many
experiments~\cite{PhysRevLett.49.91,PhysRevLett.75.4337,PhysRevA.57.3229,
  PhysRevLett.82.1345} which are Bohm's modifications of the
Einstein-Podolsky-Rosen (EPR) famous thought
experiment~\cite{PhysRev.47.777}.  As a consequence of the ingenious
paper by John S.  Bell~\cite{Physics.1.195}, we know that the
Einstein's idea to find a local realistic theory describing quantum
phenomena~\cite{AEinstein.P-S} was false: the nature of quantum world
is non-local.  This is related to the property of multiparticle states
known as quantum entanglement.  Although quantum non-locality is not
in a conflict with the Galilean physics, it is rather baffling from
the special relativity (SR) point of view.  According to
Shimony~\cite{EventProcesQuantWorld} quantum mechanics and special
relativity might ``peacefully coexist'' if they attribute physical
meaning to the final probabilities only.  However, this ``peaceful
coexistence'' relies on the ignoring some weak points of the
relativistic QM, like lack of the covariant notion of localization or
problem with spin
observable~\cite{LocalSpaceQuantPhysic,terno:014102,caban:014102,rembielinski:80}.
It would be advisable for quantum mechanics if it admit the absolute
simultaneity notion and consequently a preferred frame of
reference~\cite{Speakable,QuantMechanCosmol} (which seems to be a
natural framework for the non-locality) while preserving the (well
established) Lorentz-covariance.  In this paper we suggest how to
verify whether a preferred frame of reference in the quantum world
exists.

It seems that the present technological progress allows experimental
investigations of the relativistic aspects of quantum entanglement.
However, in description of the relativistic version of the EPR-Bohm
experiment it is necessary to overcome very serious interpretational
and theoretical difficulties (e.g.\ problem of the covariant
localization as well as the non-uniqueness and non-covariance of the
relativistic spin operator).  Another issue is that non-local
(instantaneous) reduction of quantum entangled state (if it has a
physical meaning) intuitively conflicts with relativity of
simultaneity for moving observers
\cite{PhysRevD.24.359,PhysRevLett.68.2981}.  Hardy's gedanken
experiment suggests that every realistic quantum theory should
posseses an absolute notion of simultaneity, or equivalently a
preferred frame (PF) of reference~\cite{PhysRevLett.68.2981}.  It is
worth to stress that also Bell was convinced that a consistent
formulation of quantum mechanics requires a preferred frame at the
fundamental level as the most natural way to incorporate quantum
non-locality:
\begin{quote}
  ``\ldots The aspects of quantum mechanics demanding non-locality
  remain in relativistic quantum mechanics.  It may well be that a
  relativistic version of the theory, while Lorentz invariant and
  local at the observational level, may be necessarily non-local and
  with a preferred frame (or aether) at the fundamental level\ldots''
  \cite{QuantMechanCosmol}
\end{quote} 
(J. S. Bell gives the very clear point of view on this question in
\cite{GhostAtom} too).  Similar statements were formulated by
D.~Bohm~\cite{GhostAtom,PhysRev.85.166}.  The hypothesis of the
privileged frame was also discussed by P. A. M.
Dirac~\cite{Nature:168.906}.

In this paper, we show that verification whether a preferred frame of
reference in the quantum world exists can be done by experiment of
Einstein-Podolsky-Rosen type with relativistic massive entangled pair
of spin one-half or spin one particles.  In the discussion of the
possible experiments we used the recently found relativistic EPR
correlation functions \cite{PhysRevA.77.012103,caban:014102}.  Our
analysis suggests that it is enough to perform the experiment by a
pair of observers staying in the same inertial frame.  Moreover, even
in the case of the negative answer, this experiment can give us hint
about the proper choice of relativistic spin operator.

\section{EPR correlations in preferred frame quantum mechanics}
\label{sec:PFQM}

Motivated by the above ideas, in the papers
\cite{PhysRevA.59.4187,PhysRevA.66.052114} Lorentz-covariant quantum
mechanics with a preferred frame built in was formulated.  The crucial
point in this construction is some freedom in formulation of SR, known
as the conventionality of distant clocks synchronization
\cite{AxiomTheorRelat,SomeFoundProblSpecialTheorRelat,PhysRevD.45.403,%
  PhysRep.295.93,AnnPhys.14.71}.  It is related to the fact that only
round-trip velocity of light is testable without prior synchronization
of distant clocks.  In that scheme relativity principle may be broken
without destroying Lorentz covariance.  Quantum theory formulated in
different synchronization schemes is unitary non-equivalent to the
standard relativistic quantum mechanics and leads to different
physical predictions for non-local phenomena, while it gives the same
results for the local ones.  As was shown
in~\cite{PhysRevA.66.052114}, in the Lorentz-covariant preferred frame
quantum mechanics (shortly: preferred frame quantum mechanics, PFQM)
it is possible to calculate the EPR correlation function
Lorentz-covariantly and uniquely.  For the spin-$\frac{1}{2}$
particles in the singlet state the correlation function reads
\begin{equation}
  \label{eq:0}
  C(\vec{a}, \vec{b}) = - \vec{a} \cdot R^\mathrm{T}(\Lambda, \vec{u}_A,
\vec{u}_B) \vec{b}\,,
\end{equation}
where $\vec{u}_A$, $\vec{u}_B$---velocities of the PF with respect to
the observers $A$ and $B$ (say Alice and Bob), respectively;
$R(\Lambda, \vec{u}_A, \vec{u}_B)$---the corresponding Wigner rotation
parametrized by the velocities $\vec{u}_A$, $\vec{u}_B$
(see~\cite{PhysRevA.66.052114}).  In particular, in the case when the
velocities $\vec{u}_A$ and $\vec{u}_B$ are small, i.e.\
$|\vec{u}_{A,B}/c|\ll 1$ the correlation function for two spin
one-half particles in the singlet state is given by the following
approximate formula \cite{PhysRevA.66.052114}
\begin{equation}
  \label{eq:1}
  C(\vec{a}, \vec{b}) \simeq -\vec{a} \cdot \vec{b} 
  - \frac{(\vec{a} \times \vec{b}) \cdot (\vec{u}_A \times \vec{u}_B)}{2}\,.
\end{equation}
Hereafter we put $c = 1$ and as usual $\vec{a}$ and $\vec{b}$ are
directions of spin projections measured by Alice and Bob.  If the PF
is identified with the cosmic background radiation frame then for the
observers on the Earth $|\vec{u}_{A,B}| \approx 0.001$ and the
correction to the non-relativistic formula is of the order $10^{-6}$.
It is important to stress that in the case when Alice and Bob are at
rest in the same inertial frame ($\vec{u}_A = \vec{u}_B$, $\Lambda = I$), this
correlation function takes exactly non-relativistic form
irrespectively of the PF speed \cite{PhysRevA.66.052114}, i.e.\
regardless of the choice of this inertial frame:
\begin{equation}
  \label{eq:2}
  C(\vec{a}, \vec{b}) = -\vec{a} \cdot \vec{b}\,.
\end{equation}
The same holds for correlations of spin one particles in a singlet
state; the correlation function in this case is given again by the
Eq.~(\ref{eq:2}) (with the additional factor $2/3$ in the front). Thus
for observers \emph{in the same inertial frame} PFQM predictions in
these cases are identical as in the non-relativistic QM.

\section{Relativistic EPR correlations for spin one-half particles}
\label{sec:SRQM-half}

Completely different situation takes place in the standard
relativistic quantum mechanics. Firstly, there is a problem with the
choice of the spin operator.  Indeed, spin is defined as a difference
of the total angular momentum (generator of rotations) and the orbital
angular momentum $\vec{Q} \times \vec{P}$ where the position operator
$\vec{Q}$ is non-covariant and non-unique.  If it is taken as the
famous Newton--Wigner position operator
\cite{RevModPhys.21.400,LocalSpaceQuantPhysic} then, for two EPR spin
one-half particles with velocities $\vec{v}_A$ and $\vec{v}_B$ (so
being in sharp momentum states) and mass $m$, the correlation
function in the Lorentz singlet state, for observers staying in the
same inertial frame, can be derived from the formula obtained in
\cite{caban:012103,caban:042103} and reads:
\begin{multline}
  \label{eq:3}
  \lefteqn{C^{1/2}_\mathrm{NW}(\vec{a}, \vec{b}) 
  = -\vec{a} \cdot \vec{b} 
  + \frac{(\vec{v}_A \times \vec{v}_B)}{1 - \vec{v}_A \cdot
    \vec{v}_B + \sqrt{(1 - \vec{v}_A^2) (1 - \vec{v}_B^2)}}} \\
  \cdot \left[\vec{a} \times
      \vec{b} + \frac{\vec{a} \cdot \vec{v}_A (\vec{b} \times
        \vec{v}_B) - \vec{b} \cdot \vec{v}_B (\vec{a} \times
        \vec{v}_A)}{\left(1 + \sqrt{1 - \vec{v}_A^2}\right) \left(1 +
          \sqrt{1 - \vec{v}_B^2}\right)}\right] \,.
\end{multline}
Notice that in the center-of-mass frame of EPR particles (i.e.\ for
$\vec{v}_A = -\vec{v}_B$) the correlation function takes its
non-relativistic form~\eqref{eq:2}.

In the case when we choose the so called center-of-mass (c.m.)
position
operator~\cite{LocalSpaceQuantPhysic,schroedinger:30,pryce:48,fleming:65}
we obtain an alternative spin operator used by
Czachor~\cite{PhysRevA.55.72}.  The corresponding correlation function
$C^{1/2}_\ab{c.m.}(\vec{a}, \vec{b})$ can be derived
from~\cite{caban:014102}
\begin{widetext}
\begin{multline}
  \label{eq:3.5}
  C^{1/2}_\ab{c.m.}(\vec{a}, \vec{b}) = \frac{1}{\sqrt{1 - \vec{v}_A^2
      + (\vec{a} \cdot \vec{v}_A)^2} \sqrt{1 - \vec{v}_B^2 + (\vec{b}
      \cdot \vec{v}_B)^2}} \left\{-\vec{a} \cdot \vec{b} \sqrt{1 -
      \vec{v}_A^2} \sqrt{1 - \vec{v}_B^2} + (\vec{a} \cdot \vec{v}_A)
    (\vec{b} \cdot \vec{v}_B) \vphantom{\left[1 - \vec{v}_A \cdot
        \vec{v}_B + \sqrt{1 - \vec{v}_A^2}
        \sqrt{1 - \vec{v}_B^2}\right]^{-1}} \right.\\
  \left.- \frac{\left[\vec{a} \cdot \left(\vec{v}_A \sqrt{1 -
            \vec{v}_B^2} + \vec{v}_B \sqrt{1 -
            \vec{v}_A^2}\right)\right] \left[\vec{b} \cdot
        \left(\vec{v}_A \sqrt{1 - \vec{v}_B^2} + \vec{v}_B \sqrt{1 -
            \vec{v}_A^2}\right)\right]}{1 - \vec{v}_A \cdot \vec{v}_B
      + \sqrt{1 - \vec{v}_A^2} \sqrt{1 - \vec{v}_B^2}}\right\}
\end{multline}
\end{widetext}
\begin{floatequation}
\mbox{\textit{see eq.~\eqref{eq:3.5}}}
\end{floatequation}
\addtocounter{equation}{-1}

Notice that for $\vec{v}_A = -\vec{v}_B$ this formula does not take
the non-relativistic form~(\ref{eq:2}).

\begin{figure}[ht]
  \onefigure[width=\columnwidth]{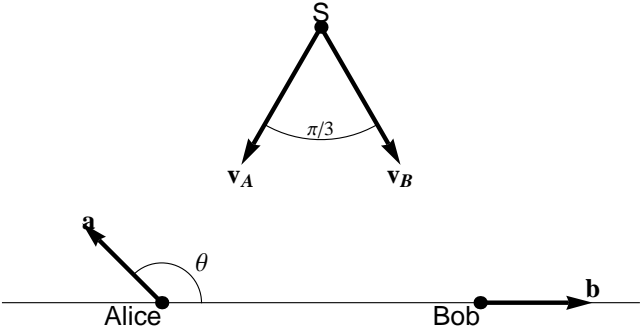}
  \caption{Alice and Bob staying in the same inertial frame perform
    the EPR experiment with the spin one-half particles.  They choose
    the coplanar configuration with $\vec{v}_A = v (-1/2,
    -\sqrt3/2,0)$, $\vec{v}_B = v (1/2, -\sqrt3/2,0)$, $\vec{a} =
    (\cos\theta, \sin\theta, 0)$, $\vec{b} = (1, 0, 0)$.  Here $S$
    denotes the EPR source.}
  \label{fig:2}
\end{figure}

\begin{figure}[htb]
  \onefigure[width=\columnwidth]{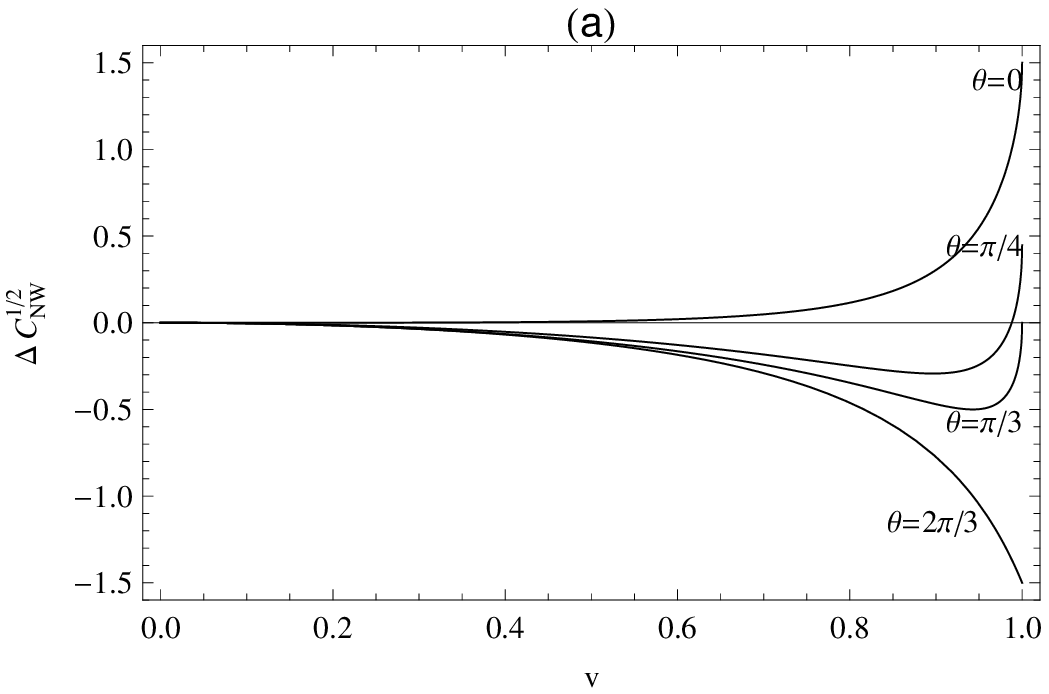}
  \onefigure[width=\columnwidth]{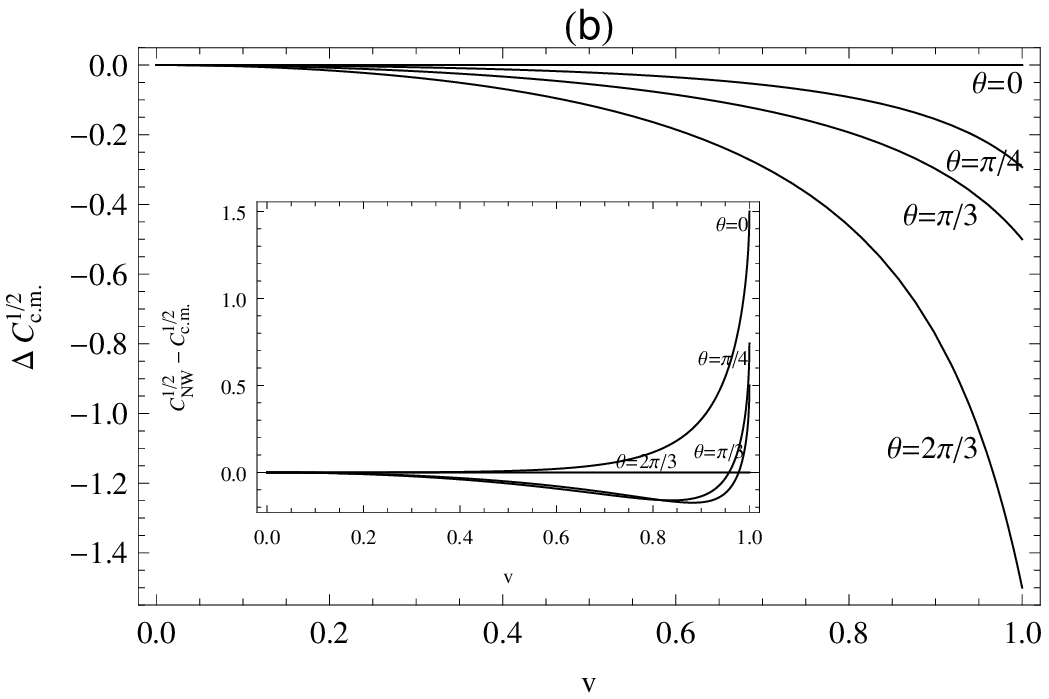}
  \caption{Difference $\Delta C^{1/2}_\mathrm{NW}$ (a) and $\Delta
    C^{1/2}_\ab{c.m.}$ (b) between relativistic and preferred frame
    correlations for the spin one-half EPR particle as the function of
    the velocity $v$ given for the angle $\theta = 0, \frac{\pi}{4},
    \frac{\pi}{3}, \frac{2 \pi}{3}$, as well as the difference between
    both approaches to relativistic correlation functions (b).}
  \label{fig:4}
\end{figure}

{} From the Eqs.~(\ref{eq:3}) and~(\ref{eq:3.5}) it follows that the
relativistic corrections to the non-relativistic (or PFQM)
correlations are of the order of $\vec{v}^2$, where $\vec{v}$ is the
average particle velocity given in the units $c$. In a specific
coplanar configuration presented in the Figure~\ref{fig:2},
where $|\vec{v}_A|=|\vec{v}_B|=v$, the difference $\Delta
C^{1/2}_\ab{NW,c.m.}  = C^{1/2}_\ab{NW,c.m.}(\vec{a}, \vec{b}) -
(-\vec{a} \cdot \vec{b})$ between relativistic and PF correlations as
the function of $v$ (in the units $c$) for selected values of the
angle $\theta$ is shown in the Figure~\ref{fig:4}.

\begin{figure}[ht]
  \onefigure[width=\columnwidth]{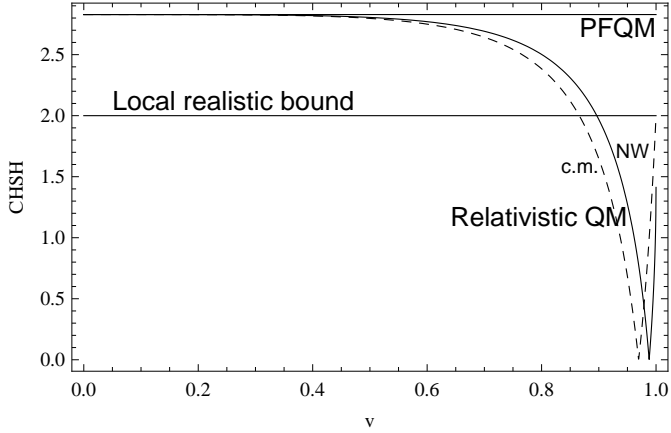}
  \caption{CHSH inequality for spin one-half particles in the
    configuration in which it is maximally broken in the PFQM compared
    to the relativistic QM predictions with Newton--Wigner (solid) and
    center-of-mass (dashed) spin operator.}
  \label{fig:10}
\end{figure}

Now, let us consider the Clauser--Horne--Shimony--Holt inequality
\begin{equation}
  \label{eq:5}
  \mathrm{CHSH} = |C(\vec{a},\vec{b}) - C(\vec{a},\vec{d}) 
  + C(\vec{c},\vec{b}) + C(\vec{c},\vec{d})| \leq 2\,,
\end{equation}
where Alice measures spin projections along the directions $\vec{a}$
and $\vec{c}$ while Bob along the directions $\vec{b}$ and $\vec{d}$.
This inequality is maximally violated by preferred frame quantum
correlations for, \emph{e.g.}, $\vec{a} = (-1/\sqrt2, 1/\sqrt2,0)$,
$\vec{c} = (1/\sqrt2, 1/\sqrt2,0)$, $\vec{b} = (0, 1, 0)$ and $\vec{d}
= (1, 0, 0)$, i.e.\ when $\mathrm{CHSH} = 2\sqrt2$.  In the
Figure~\ref{fig:10}
we have shown both PF and relativistic CHSH as functions of the
particle velocity.

Summarizing, in all considered situations with spin one-half EPR
particles we observe for ultra-relativistic velocities significant
differences between predictions of relativistic and preferred frame
quantum mechanics.  Moreover, there are also significant differences
between correlation functions $C^{1/2}_\mathrm{NW}$ and
$C^{1/2}_\ab{c.m.}$ (see Fig.~\ref{fig:4}(b) and Fig.~\ref{fig:10}).

\section{Relativistic EPR correlations for spin one particles}
\label{sec:SRQM-one}

In the case of the spin one particles, the relativistic EPR
correlations differ from the PFQM ones also in the center of mass
frame of reference, i.e.\ for $\vec{v}_A = -\vec{v}_B \equiv \vec{v}$.
Indeed, from the paper \cite{PhysRevA.77.012103} it can be deduced
that in this case the relativistic correlation functions are of the
form
\begin{align}
  \label{eq:4}
  C^1_\mathrm{NW}(\vec{a}, \vec{b}) &= \frac{2 (1 - \vec{v}^2)}{3 - 2
    \vec{v}^2 + 3 \vec{v}^4} \nonumber\\
  &\quad \times \left[-\vec{a} \cdot \vec{b} (1 + \vec{v}^2)
    + 2 (\vec{a} \cdot \vec{v}) (\vec{b} \cdot \vec{v})\right]\,,\\
  \label{eq:4.5}
  C^1_\ab{c.m.}(\vec{a}, \vec{b}) &= \frac{2 (1 - \vec{v}^2)^2}{3 - 2
    \vec{v}^2 + 3 \vec{v}^4} \nonumber\\
  &\quad \times \frac{-\vec{a} \cdot \vec{b} (1 + \vec{v}^2) +
    (\vec{a} \cdot \vec{v}) (\vec{b} \cdot \vec{v})}{\sqrt{\left(1 -
        \vec{v}^2 + (\vec{a} \cdot \vec{v})^2\right) \left(1 -
        \vec{v}^2 + (\vec{b} \cdot \vec{v})^2\right)}}\,.
\end{align} 

\begin{figure}[ht]
  \onefigure[width=\columnwidth]{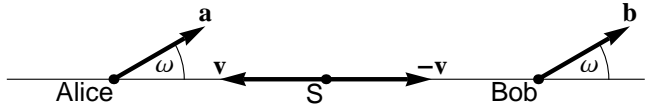}
  \caption{Alice and Bob staying in the same inertial frame perform
    the EPR experiment with the spin one particles.  They choose the
    coplanar configuration with $\vec{v}_A = v (-1,0,0)$, $\vec{v}_B =
    v (1,0,0)$, $\vec{a} = (\cos\omega,\sin\omega,0) = \vec{b}$.}
  \label{fig:5}
\end{figure}

We observe a maximal deviation from the PF correlations in the
coplanar configuration with $\vec{a} \cdot \vec{b} = 1$ (or $-1$),
$\vec{a} \cdot \vec{v}/|\vec{v}| = \vec{b} \cdot \vec{v}/|\vec{v}| =
\cos\omega$ (see Figure~\ref{fig:5}).

\begin{figure}[ht]
  \onefigure[width=\columnwidth]{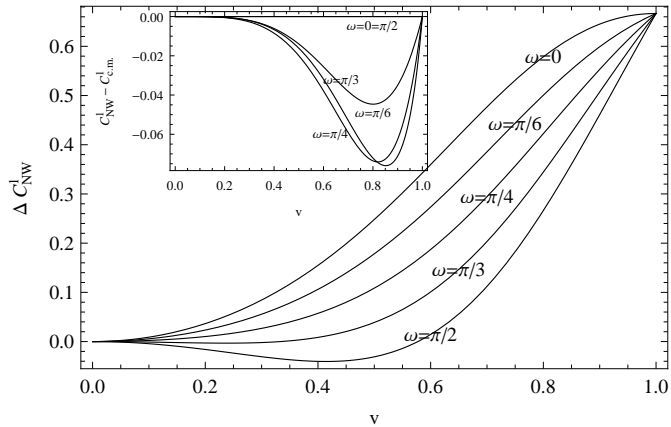}
  \caption{Difference $\Delta C^1_\mathrm{NW}$ between relativistic
    and preferred frame correlations for spin one EPR particles as the
    function of the velocity $v$ for given selected values of the
    angle $\omega$, as well as the difference between
    $C^1_\mathrm{NW}$ and $C^1_\ab{c.m.}$.}
  \label{fig:7}
\end{figure}

In the Figure~\ref{fig:7}
we have shown the dependence of $\Delta C^1_\mathrm{NW}$ on velocity
for selected values of the angle $\omega$, where the difference
$\Delta C^1_\mathrm{NW,c.m.} =
C^1_\mathrm{NW,c.m.}(\vec{a}, \vec{b}) - (-2 \vec{a} \cdot
\vec{b}/3)$.

\begin{figure}[htb]
  \onefigure[width=\columnwidth]{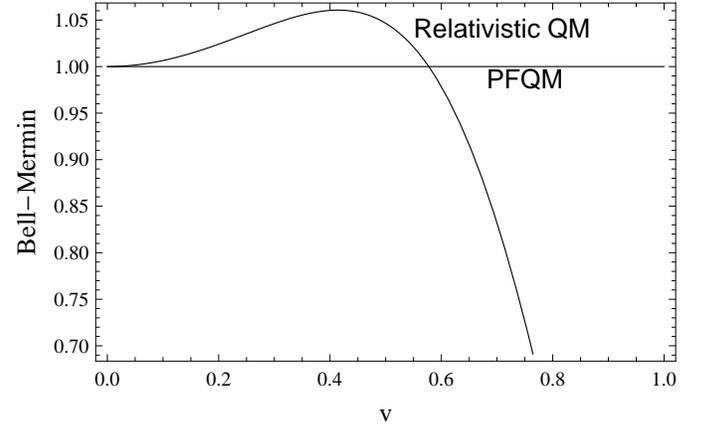}
  \caption{Breaking of the Bell--Mermin inequality by the relativistic
    QM for the configuration with coplanar vectors $\vec{a}$,
    $\vec{b}$ and $\vec{c}$, $\vec{a}=(0,0,1)$,
    $\vec{b}=(\sqrt{3}/2,0,-1/2)$, $\vec{c}=(-\sqrt{3}/2,0,-1/2)$,
    $\vec{v}/|\vec{v}|=(0,-1,0)$ (in this configuration Bell--Mermin
    inequality has the same form for both Newton--Wigner and
    center-of-mass spin operators).}
  \label{fig:9}
\end{figure}

We can also observe a discrepancy between predictions of the
relativistic and preferred frame quantum mechanics in the case of
Bell-type inequalities given in the Mermin's paper
\cite{PhysRevD.22.356} for particles with spin one in a singlet state.
According to this paper, in every local realistic theory the following
inequality must be satisfied:
\begin{equation}
  \text{Bell-Mermin} = C^1(\vec{a}, \vec{b}) + C^1(\vec{b}, \vec{c}) 
  + C^1(\vec{c}, \vec{a}) \leq 1\,.
\end{equation}
One can show that in the PFQM case as well as for non-relativistic QM
this inequality holds for each configuration of $\vec{a}$, $\vec{b}$
and $\vec{c}$.  However, relativistic correlation functions
(\ref{eq:4}), (\ref{eq:4.5}) can violate Bell-Mermin inequality.  
Indeed, in the Figure~\ref{fig:9}
we observe that the standard relativistic QM breaks this inequality
for a particular configuration, while PFQM does not.

\section{Summary}
\label{sec:summary}

{} From the above discussion it is clear that to test the hypothesis
of existence of the quantum preferred frame it is enough to perform an
EPR type experiment with entangled relativistic massive particles by
observers staying in the same inertial frame. If they observe no
deviation from the non-relativistic EPR correlations and/or Bell-type
inequalities, then there is a serious indication that the quantum
preferred frame does exist.  However, even in the opposite case, the
experiment can be helpful in deciding which choice of relativistic
spin operator is correct.  To our best knowledge no such experiment
has been yet performed---it must involve ultra-relativistic massive
spin one-half or spin one particles.  This excludes any experiment
with photons as well as with kaons (or $B$ mesons) or non-relativistic
heavy ions and/or atoms.  A serious advantage of using massive
fermions to test Bell-type inequalities lies in the fact that these
particles, contrary to the photons, are well localized (its coherence
length is of the order of $\unit{10^{-15}}{\metre}$, while for photons
$\sim\unit{1}{\metre}$).  Moreover, the singlet state of the EPR pair
is well defined by measuring the internal energy of the system.

 As far as we know, up to date there has been only three experiments
 testing Bell-type inequalities by means of massive relativistic spin
 one-half particles: the Lamehi-Rachti--Mittig (LRM) experiment
 \cite{PhysRevD.14.2543} performed about thirty years ago in
 CEN-Saclay and two recent experiments: the first one given at the
 Kernfysisch Versneller Instituut (KVI, Holland) by S.~Hamieh et al.\
 \cite{hamieh04} and the second one performed by the H.~Sakai et al.\
 \cite{sakai:150405} in RIKEN Accelerator Research Facility (Japan).
 In all three experiments the proton-proton spin correlations was
 measured.  LRM team tested Bell's inequalities with use of the low
 energy ($\unit{13.5}{\mega\electronvolt}$) proton beam which
 corresponds to the proton velocity $v \sim 0.17 c$.  For the other
 hand, in KVI experiment, the spin correlations of proton pairs in a
 ${}^1S_0$ intermediate state, obtained from
 ${}^{12}\mathrm{C}(d,{}^2\mathrm{He}){}^{12}\mathrm{B}$ nuclear
 charge-exchange reaction, were measured for protons with the kinetic
 energy $\sim\unit{86}{\mega\electronvolt}$ ($v \sim 0.4 c$). Finally,
 in the RIKEN experiment the proton pair was created in the
 ${}^1\mathrm{H}(d,{}^2\mathrm{He})n$ charge-exchange reaction with
 the proton energy $\sim\unit{135}{\mega\electronvolt}$ ($v \sim 0.5
 c$). All these experiments were in agreement with the
 non-relativistic quantum mechanics predictions. However, the
 particles were too slow to give a significant difference between
 predictions of relativistic QM and PFQM.  From our estimation, it is
 clear, that in a conclusive experiment the proton energy should be
 larger than $\unit{800}{\mega\electronvolt}$ ($v \sim 0.85 c$).
In general, a correlation experiments with EPR particles with kinetic
energies of the order of their masses are waiting to settle the posed
question.

\acknowledgments One of the Authors (JR) is greateful for fruitful
discussions to Marek Czachor, Ryszard Horodecki, Reinhard Werner and
Marek \.Zukowski.  We also would like to acknowledge Hideyuki Sakai
for his explanation of the experiment performed by his group.  This
work has been supported by the University of \L\'od\'z grant and by
LFPPI network.


\end{document}